# Game Development-Based Learning Experience: Gender Differences in Game Design

Bernadette Spieler, Wolfgang Slany
Graz University of Technology, Graz, Austria
bernadette.spieler@ist.tugraz.at
wolfgang.slany@tugraz.at

**Abstract:** Learning theories emphasize the importance of intrinsic and extrinsic motivators in curricula, and games are a promising way to provide both while constructing the game and presenting or sharing it in public or with a community. New technologies and the emerging mobile gaming sector further the case that learning should be promoted everywhere and anytime. What seems to be a promising opportunity for all pupils to learn coding in an entertaining way raises the question of whether such game based concepts also help to fix the gender gap of women in IT related fields. Gender differences are already present in secondary schools. These are the years where first career choices but also low levels of participation in technical subjects occur. To address this gender bias, a goal of the European project No One Left Behind (NOLB) was to integrate Pocket Code, an app developed by Catrobat, a free open source non-profit project at the University of Technology in Graz/Austria, into different school subjects, thus making coding more accessible and attractive to female pupils. During the period of this project (2015-2017), teachers were supported to guide their pupils in the learning processes by constructing ideas and realizing them through game design. Furthermore, teachers were trained to teach good game design and game production by sticking to an agile game design approach (research, design, development, testing, sharing), and the Mechanics, Dynamics, and Aesthetics framework, as well as other techniques, like the use of storyboards, and "ceremonies". This is especially important for girls; according to the literature, they are less likely to play games. For this paper an analysis of submitted programs according to their game design has been performed. In detail, the evaluation considered formal game elements, gaming structures, and used graphics, as well as Pocket Code specific aspects. The program's analysis showed commonly used design patterns by genders and suggests preferred game design characteristics of female teenagers. This analysis helps to build a more creative and inclusive Game Development-Based Learning (GDBL) environment that provides room for self-expression and inspires by building on intrinsic and extrinsic motivators by constructing personal experiences.

**Keywords:** Pocket Code, game design, gender design, gender, mobile learning, coding

## 1. Introduction

Much effort can be recognized from the Austrian government (Federal Ministry of Education Austria, 2017), the European Commission (EC, 2016), and from institutions all over the world (Code.org, 2018; Ladner and Israel, 2016) to apply Game-Based Learning concepts (GBL or Game Development-Based Learning/GDBL) from primary school to higher education. In addition, games are known as an effective approach for teachers to motivate pupils to interact and communicate as well as to learn (Kafai and Vasudevan, 2015). To play an active part in the learning process and to create something meaningful allows pupils to collaborate and construct solutions for problems. Uptakes of the constructionist approach (Papert and Harel, 1991) to foster playful activities in teaching are increasing within the education communities, e.g., K-12 movements in the US (Kopcha, Ding, and Neumann, 2016). Through an interdisciplinary approach, teachers not only transfer their subject knowledge but also teach fundamental programming skills. Playful coding activities are therefore a perfect match of development of creativity, problem solving, logical thinking, system design, and collaboration skills (Backlund and Hendrix, 2013). In that way, pupils can learn about a specific concept or subject by developing personalized games, allowing pupils to freely select and use content-related preferences, like genres, themes, goals, characters, game dynamics and mechanics, backgrounds, or assets.

This paper is organized as follows: First, the paper provides a literature review by presenting intrinsic/extrinsic motivators, gaming elements, game design principles, and game-based learning theories. This section concludes by focusing on games and gender issues. This should suggest optimal conditions from the literature on how to use GDBL in an academic setting. Subsequently, Section 3 presents the app and the related NOLB project. Section 4 present the results of the game design analysis by gender and Section 5 and 6 concludes this paper and describes the authors' future work.

## 2. Games and Learning

Engaging activities are well-received by pupils and contribute to improving their overall motivation and productivity (Khaleel and Meriam, 2015). Within suitable game-based learning environments, learners are encouraged to be creative and collaborate with others. In addition, learners acquire conceptual understanding as well as practical skills (Singer and Schneider, 2012). Hence, the use of digital games as part of the formal academic curriculum comes as a natural response (Kerr, 2006). The concepts of intrinsic motivation and the flow of experiences are best applied if pupils are freely engaging in self-chosen activities with a playful twist (Brophy, 2013).

The literature distinguishes between two types of reinforcement: external and internal reinforcement (Ryan and Deci, 2000). Only the external is perceived directly. Both encourage people to repeat their previous behavior. External reinforcement provides motivation, but often the learner sets the wrong goals. Examples for external motivators are the earning of certificates or awards. Therefore, extrinsic motivation is associated with Skinner's behavioral theories of human learning (Skinner, 1976). In contrast, internal motivation builds on praising pupils and recognition or acknowledgements (Hamari and Kovisto, 2013). This gives the pupils the feeling of pride and achievement and supports self-directed learning and practical relevance.

### 2.1 Game Design Elements and the Game Design Process

Most characteristics are very similar in all games. For instance, rules, goals, and uncertain outcomes (Seaborn and Fels, 2015). Games are formed from a variety of components and it is the players' perception, which determines whether the experience is fun and entertaining or not. In addition, the Mechanics, Dynamics, and Aesthetics (MDA) of a game motivates the players to keep on playing and to be successful (Ibáñez et al, 2014). Different dimensions exist within games that characterize them (Arseth, 2003). The game design elements can be broken into three subcategories:

- Gaming-world (e.g., level design, theme, genres, like adventure, action, puzzle, simulation, strategy)
- Game-structure (the rules of the game and the goal, MDAs)
- Game-play (e.g., the story, the player and their actions, strategies, and motives)

The description of game genres and MDAs was already part of the authors' previous work (Spieler et al, 2017). In short, *Mechanics* describe the particular components of the game and the constructs of rules or methods for the gameplay (e.g., points, level, or challenges). *Dynamics* are defined through interactions of users with the game mechanics and describe the play of the game when the rules are set in motion (e.g., rewards, status, or achievement). *Aesthetics* refer the player's experience with the game (e.g., fantasy, narrative, or discovery).

To transfer the concept of games in a consistent structure, several strategies are possible. For the European project NOLB, the team used the term "ceremony" or "Shape of a Game" to identify the different "scenes" of a game (title, introduction, levels, end screen). This shape of the interface should support overall clarity, help to focus on the relevant elements of the game, and follow a clear method of interaction (Martinovs et al, 2017).

For game development cycles, concepts of agile and iterative software development can be used to leverage this process and to see first results very quickly (DeMarco-Brown, 2013). As the scope of this paper is limited, only a simplified life cycle is visualized to describe the process of how a team works in iterations to deliver or release software:

- *Research*: story, title, genre, and theme of the game is selected, as well as a rough concept about the structure and gameplay
- *Design*: the artwork, game content and other elements (characters, assets, avatars, etc.) are produced
- *Development*: includes all of the actual programming, followed by *Testing* the code
- *Release*: as a final step

This concept should help, before concentrating on every detail of the game, focus on the smallest step the games needs for playing it, e.g., limited interface control but basic game functionality (Salen and Zimmerman, 2003).

### 2.2 Game-Based Approaches for Fearning

Serious games, as an overall term and involve all aspects of education, e.g., teaching, training, and informing (Michael and Chen, 2006). GBL is the practice of creating serious games to improve teaching activities and ini-tiatives by virtue of its factors concerning engagement, motivation, role-playing, and repeatability (Backlund and Hendrix, 2013). The goal of Gamification is to apply game elements in a non-gaming context (Glover, 2013). Thus, Game Development-Based Learning (GDBL), is defined as learning by making games, includes all these aspects (see Figure 1) (Wu and Wang, 2012).

In this constructionist-learning environment, learners are encouraged to develop their own games and make design decisions (Kafai, 2006). Benefits of designing games include enhancing problem-solving skills, fostering of creativity and self-directed learning, and learning how to code in a fun way. The game construction process encourages further collaboration and engagement through teamwork within the classrooms. Therefore, creating personalized games is a popular approach to introduce novice users to programming, to conceptual thinking, engineering, art, and design. In this context, not only can a game be used for learning, but also the game development tools be used for studying relevant topics within computer science, software engineering (SE), or game programming through motivating assignments.

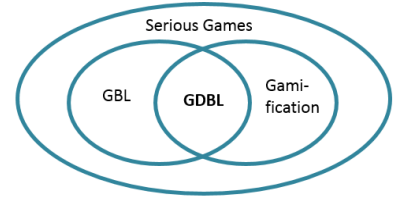

**Figure 1:** Relationship of Serious Games, GBL, Gamification, and GDBL

### 2.3 Games and Gender

A closer look on girls' typical design patterns has been already presented in the author's previous work (Spieler, 2018). According to a study (Google, 2017), a percentage of 43% of women play mobile games five times per week or more (38% of men to so). This study also mentions that even the Google Play Store on its own, features more male identified characters than female (44% more male characters than female in the 100 most revenue-generating games). In addition, only 30% of the women who play feel that the games are made for them. Google believes that mobile gaming can inspire creativity, show users new worlds, and let them engage and they sees a huge opportunity to make mobile games more diverse, inclusive, and engaging for all players (Bevans, 2017). Current numbers of the video game industry showing that 46% of gamers across 13 countries are women (aged between 10-65) and that the favorite game genres among different consoles are action/adventure, strategy, or puzzle games (NewZoo, 1017). This indicates games of different genres, characters, or themes that were commonly played more by females. A closer look at the developed games during the NOLB project (see next section) should show if this is also the case for the games that are developed by females.

### 3. Pocket Code and the No One Left Behind (NOLB) Project

The European No One Left Behind[1] (NOLB) project has been funded by the Horizon 2020 framework, involved partners from Germany, Spain, the UK, and Austria and evaluated its results from January 2015 to June 2017. The vision of the NOLB project was to unlock inclusive gaming creation/game design and to construct experiences in formal learning situations through secondary level, particularly for teenagers at risk of social exclusion. The study in Austrian was seen as a chance to recognize gender differences in engaging with coding and game design activities. Thus, the app Pocket Code[2], a media-rich programming environment for teenagers to learn coding with a visual programming language very similar to the Scratch[3] environment, has been used for creating games and apps. Pocket Code is freely available at the Google Play Store and allows the creation of games, stories, animations, and many types of other apps directly on phones, thereby teaching fundamental programming skills. During NOLB, pupils used Pocket Code for creating small classroom projects in different subjects like, fine arts, computer science, physics, music, or English. This app is developed at the University of Technology in Graz/Austria (TU Graz) as part of the Free Open Source Software (FOSS) project Catrobat[4].

### 4. Game Design Analysis by Gender

Different types of game analysis exist (Arseth, 2003), e.g., players' reports, walkthroughs, discussions, observing others play, interviews with players, game documentation, playtesting reports, or interviews with game

---

[1] No One Left Behind EC project H2020: http://no1leftbehind.eu/
[2] Pocket Code Google Play Store: https://catrob.at/pc
3 Scratch MIT: https://scratch.mit.edu/
4 Catrobat: https://www.catrobat.org

developers. For this analysis, the formal elements are the most interesting (Lankoski and Björk, 2015). For the evaluation of the created Pocket Code games, different game elements, the game mechanics and the goals they relate to (MDA) have been identified. In addition, special characteristics of the programmed games (theme, genre, goal, etc.) have been analysed. In detail, the evaluation considered formal game elements, structure, and graphics of the created games (Jesper, 2007). The following analysis has been made:
- Definition of the used genre, theme, and goal of the games
- Description of the used MDA: mechanics, dynamics, and aesthetics,
- Usage of formal elements, e.g., level of control (interaction), visual design, sound design, restart/end of the game, and the "Shape of a Game" ceremony
- Other design elements: main character, side characters
- Pocket Code specific aspects, e.g., used sensors, amount of bricks, scripts, objects, looks, or variables

Altogether, 77 programs have been analysed which have been created during the course of the NOLB project in Austria by pubis Year 8 to 12 in English, computer science, and fine arts: girls created 37 of the programs, boys 34, and 8 programs were created in mixed gender teams. The number of programs made in mixed teams is too small to perform a significant analysis. Thus, games made in mixed groups are not part of this analysis. By analysing the games made by boys/girls separately a gender differentiated approaches has been used to trigger a trend. This may can be criticized because "stereotypical" forecasts are often confirmed by it (Grünberg, 2011). In our case it should build gender awareness and help in setting up a more inclusive GDBL classrooms for female pubils.

### 4.1 Genre, Theme, and the Goal of the Games

Figure 2 shows the used genres/themes of female pupils and Figure 3 games created by male pupils.

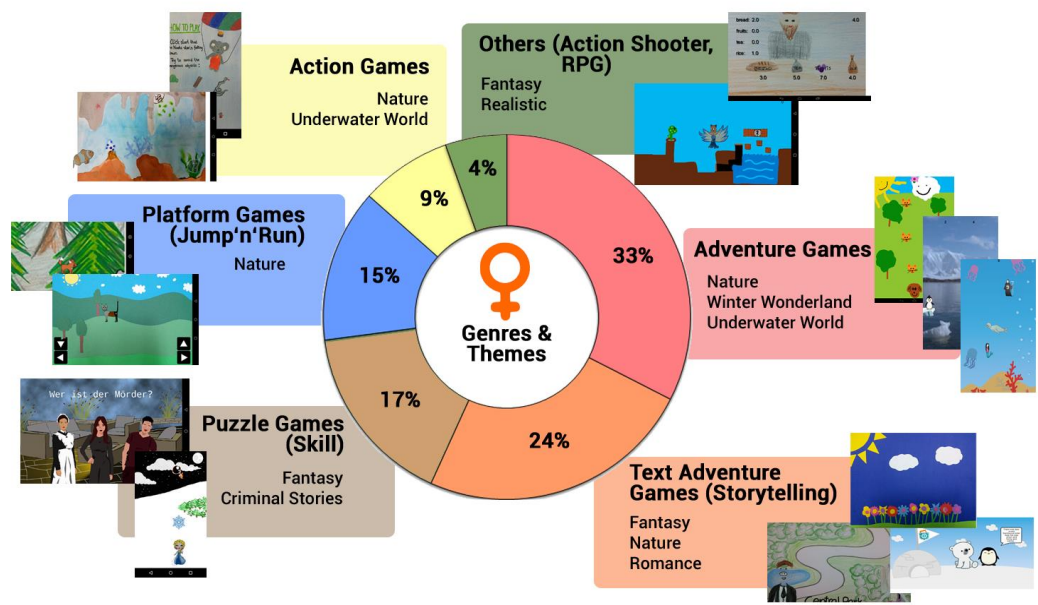

**Figure 2:** Created games by girls categorized by their used genre/theme

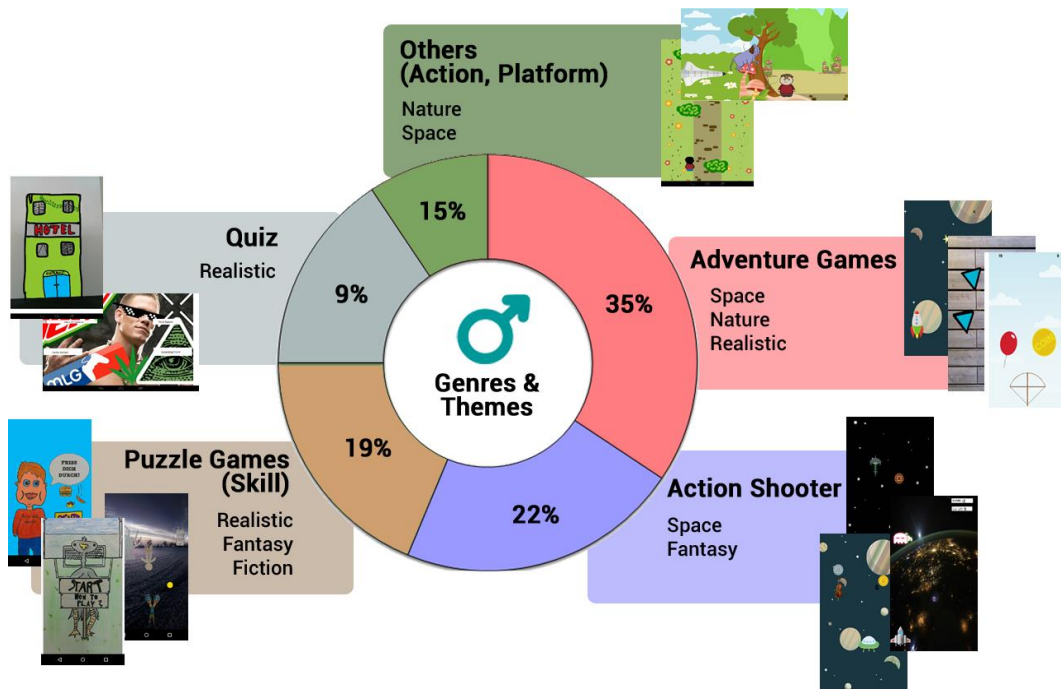

**Figure 3:** Created games by boys categorized by their used genre/theme

The games created by girls were mostly adventure games, text adventure/storytelling games, puzzle games, and platform games. None of the games created by girls falls into quiz genre. Games created by boys were also mostly adventure games, followed by action shooter games, puzzle games, quizzes, and action games. Boys, in contrast, created only two games that could be considered in the text adventure or RPG adventure genres. A Chi²-test has been performed to analyze if there is a significant relationship between the gender and the genre they choose for their games. The results indicated that the relationship between gender and genre is significant $\chi2(71)$ p = 0.008 ($\alpha$ = .05). A comparison between the two genres adventure (included also text/RPG adventure genre and action (included also shooter) showed that female pupils used significantly more often the adventure genre than male student did ($\chi2(71)$ p = .002, $\alpha$ = .05). The most popular genre among mixed gender teams was the quiz genre (n = 3). None of the created games were based on the strategy genre or the simulation genre (e.g., racing). Female pupils most often used a nature-based theme (43%) and the most popular theme among male pupils was the space theme (46%).

A closer look at the underlying goals of the games shows no significant differences between games from girls and boys. Most of the games of both genders had either the goal of capture, avoid, or destroy something (55%), collect items (11%), solve a puzzle (9%), or no goal (e.g., in storytelling games, or animations; 18%). Territorial, spatial alignment, or building something were never used as goals for any of the games. In addition, 10 of the games used the options to restart the game (12%).

### 4.2 Mechanics, Dynamics, and Aesthetics

Figure 4 provides an overview about mechanics, dynamics, and aesthetics used by gender. The size of the icons correlates with the frequency of the mentions.

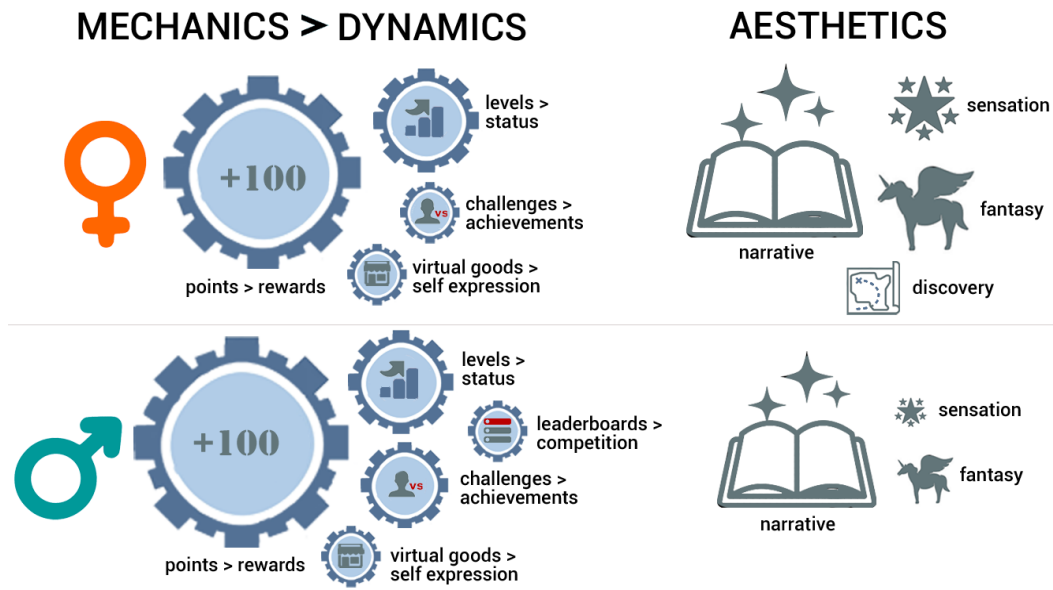

**Figure 4:** Mechanics, dynamics, and aesthetics used by gender

For mechanics and dynamics, female and male pupils both most often used points as a reward (female = 14, male = 17) and levels or lives for status (female = 7, male = 7) in their games (no significant differences). In addition, four of the games created by mixed gender teams had levels integrated. Both used virtual goods in the form of a virtual shop to purchase goods in one game, and challenges were used by male pupils five times and by female pupils once. Challenges are provided in the form of bosses or achievement of missions through levels. One program created by male pupils used leaderboards in the form of a high-score list for competition at the end of the game. In mixed groups, levels were used the most often (n = 4) for showing the status. Together, five games of mixed gender groups had both points and levels. Gifting has not be used in any of the games.

The definition of the used aesthetic elements is very objective. What one defines for him/herself as sensational could be seen as boring for another user. For the evaluation, the game has been characterized as narrative if, for instance, an intro is provided, the game has a storytelling part, or explanations are provided through the game. A number of 10 games created by girls had this property as well as seven of the games created by boys. In addition, seven of the games created by mixed gender groups are considered as narrative. Sensation is part of the game if the game surprises the player with unusual characters or new concepts (female = 7, male = 2). The game has the aesthetic of fantasy if the story includes fantasy characters or a fantasy world (female=8, male = 4). A game has the characteristics of discovery if missions lead you through a world (female = 4). Other characteristics like fellowship, expression, and submissions are difficult to integrate in a game made with the Pocket Code app and thus they were not used in pupils' games.

### 4.3 Formal Elements

In total, 40 of 77 games used the "Shape of a Game" (female = 23, male = 10). Thus, more of the half of the games had a title, instruction, and end screen integrated. Figure 5 shows which level of control has been used to control the main characters: Animation (female = 9, male = 3), inclination sensor of the device (female = 15, male = 18), buttons (female = 3), "When tapped" property (i.e., tapping on the screen or objects, female = 14, male = 10), or the use of physical properties (through the integrated physics engine, e.g., gravity, female = 1, male = 10). To conclude, both genders used mostly the inclination sensor to control their characters, or the story is continued by the "When tapped" property. The analysis of the Chi² test shows that there is a significant relation between the gender and the used level of control $\chi^2(71)$ p = 0.004 ($\alpha$ = .05).

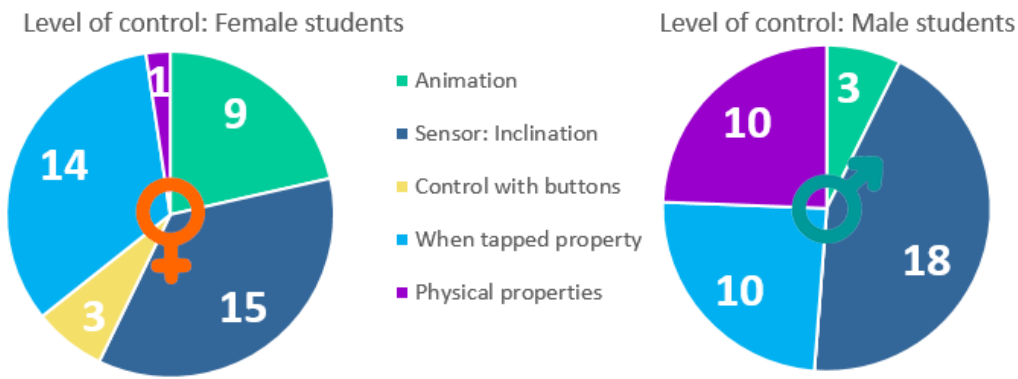

**Figure 5:** Level of control of games made by girls and boys

Another interesting question was how the graphics were created. The inspection of the programs shows that many of the imported pictures were handmade artwork which was photographed (female = 19, male = 9) or made by pupils as a whole in Pocket Paint (Pocket Paint is our second app for creating/edging objects[5], female = 8, male = 9). Male pupils mainly used pictures from the Internet for their games or utilized the predefined graphics from our media library[6]. In addition, some of the programs made by girls had self-made graphics drawn in Photoshop (n = 3). Games that have photographs integrated showed pictures of the pupils themselves. In mixed groups, the artwork was either handmade or Pocket Paint was used (n = 4). A performed Chi² test showed that overall female pupils made significantly more graphics on their own (artwork, Photoshop, Pocket Paint) than male pupils did (Internet, Media Library), $\chi 2(71)$ p = 0.030 ($\alpha$ = .05).

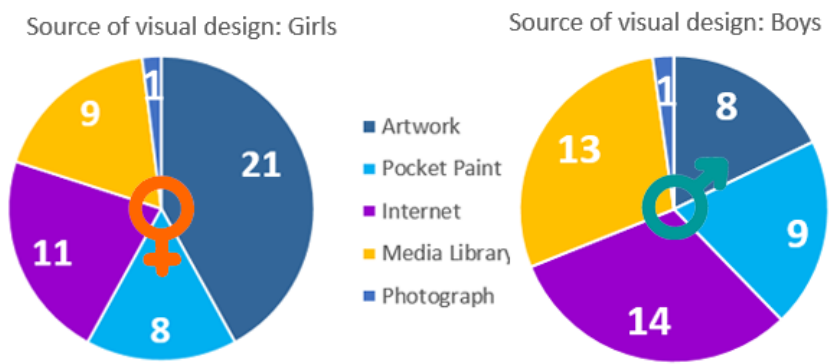

**Figure 6:** Source of visual design: boys' and girls' games

In regard to sound design, only 24 games used sounds (31%) in total. 15 games made by boys used sounds (2 used their own recordings, 5 used sounds imported from the device, and 8 imported from the media library). In games made by females, 10 sound files were found (3 used their own recordings, 1 used files imported from the device, and 6 borrowed from the media library).

### 4.4 Game Characters

A range of different graphics from different sources was used. For this evaluation, first, the main character and second, the side characters were assessed. Figure 7 shows the main characters used in games made by girls. Most games had animals as main characters, like a dog or a penguin (n = 13), or fantasy characters (n = 11), like a fairy, monsters, or a mermaid. None of the games made by female pupils had a main character which could be characterized as transportation. In contrast, games made by boys mostly consisted of main characters that represented the category of transportation (n = 12), like spaceships (n = 8), rockets (n = 3), and UFOs (n = 1). A number of six programs made by boys had male main characters. Only one game made by boys had a female character (it is a woman who balances a bucket on her head). Main characters in mixed teams were mainly male characters (n=4) or animals (e.g., a raccoon or a dog).

---

[5] Pocket Paint: http://catrob.at/PPoGP
[6] Media library for looks: https://share.catrob.at/pocketcode/media-library/looks

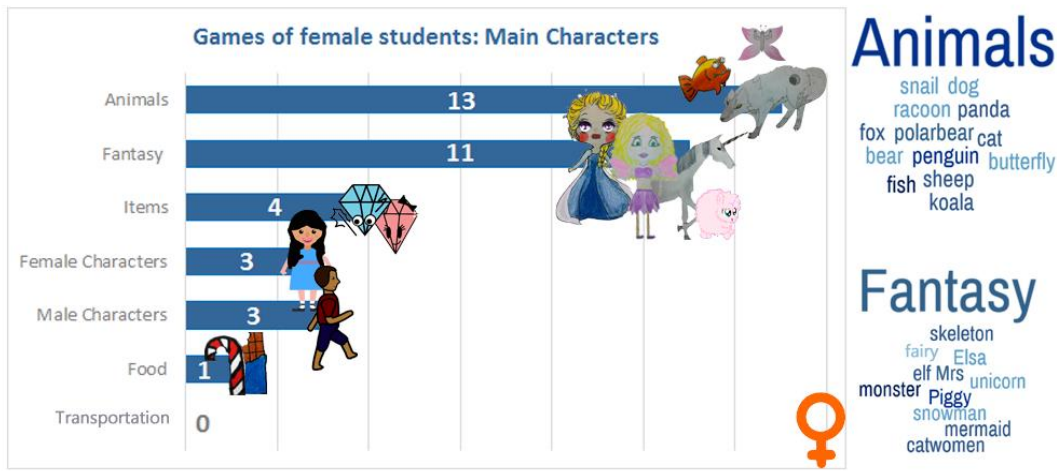

**Figure 7:** Used main characters in games made by girls

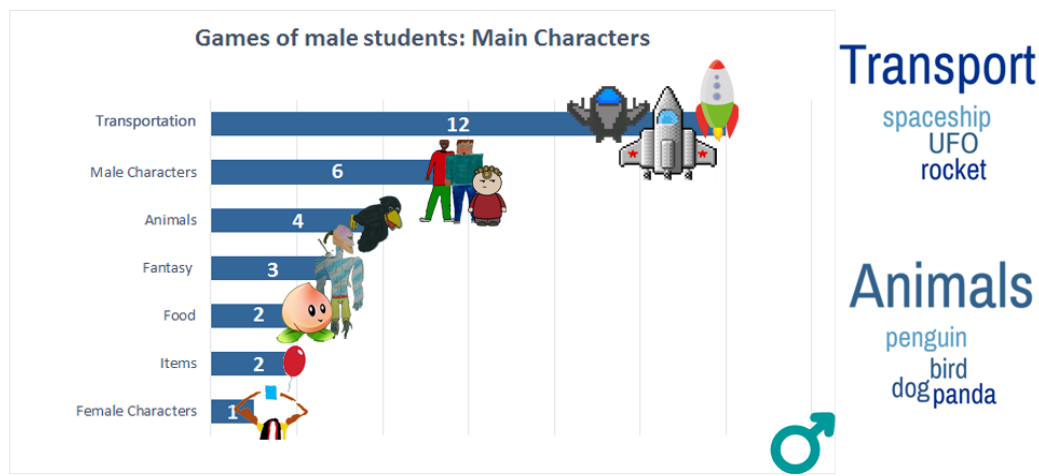

**Figure 8:** Used main characters in games made by boys

In addition, a number of side characters are used within the games. A number of 8 games made by females used animals as side characters. Examples are a seal, a cat, or a snake. In addition to fantasy characters (n = 7; e.g., angel, snowman, zombies) and many food characters were also used (n = 8, e.g., donuts, carrots, or peanuts). Side characters used by boys were mostly space stuff (n= 13), e.g., meteors, asteroids, stars, planets, or items (n = 8), e.g., a football or coins. Mixed gender groups used mostly animals (n = 3) for side characters. A wide range of different backgrounds has been used. The backgrounds used are mostly complementary with the theme of the game, thus they are not described in detail.

### 4.5 Pocket Code Specific Aspects

Finally, all created programs have been analyzed according to their number of scenes, scripts, bricks, objects, looks, sounds, and global and local used variables (see Figure 9) game that has been upload[7].The mean value for each element of all programs was calculated and analyzed by using a t-test. The results were not significant in the means of the numbers of objects in the games (p = .401, α = .05) but the mean is still interesting: female pupils used an average of 24 objects in their games, whereas male pupils used only 11 objects on average.

---

[7] Pocket Code web-share platform: https://share.catrob.at/pocketcode/

|  | female pupils | male pupils |
|---|---|---|
| number of scenes | 2,44 | 2,79 |
| number of scripts | 23,97 | 24,09 |
| number of bricks | 120,89 | 133,63 |
| number of objects | 23,89 | 10,66 |
| number of looks | 12,78 | 14,22 |
| number of sounds | 3,16 | 1,44 |
| number of global variables | 3,62 | 3,81 |
| number of local variables | 1,03 | 1,66 |

**Figure 9:** Code statistics NOLB programs

## 5. Discussion

The game design analysis shows many interesting issues regarding the different game design patterns of female and male pupils. Not only the literature stated commonly used game design patterns among genders, our female participants also liked to create significantly more games of adventure genres. In contrast, boys created significantly more shooter games and puzzle games. In addition, girls used several kinds of adventure genres (RPG, text adventure), in contrast to boys, who more frequently used the default adventure genre. The chosen themes of the games show more "stereotypical" choices between boys and girls. Girls mostly prefered nature themed graphics (woods, fields, sea, winter, sky, etc.), whereas boys chose more space, fantasy, and realistic themes which also suited the action shooter genre. Overall, female pupils spent much more time with creating their artwork. Moreover, the used MDA shows that pupils of all genders know how to integrate mechanics of points, levels, and challenges for the player, but for female pupils aesthetics are more important than for male pupils. Games created by girls used more elements of narrative, sensation, and fantasy than boys' programs.

## 6. Conclusion and Future Work

To conclude, this analysis highlights that it is important that female game designers get enough time in game design in order to create artworks and build up the concepts that drive their gaming worlds. Boys in contrast, focused less on creating sophisticated graphics and but used significantly more advanced features, like physical properties. However, Figure 9 shows that there are no significant differences in the scope of the programs. On the basis of these results, new featured games who should attract especially our female target group has been developed for the community webshare as well as new assets and templates for our app (Spieler et al, 2017). Further, the setting for our future workshops will consider the results, e.g., by providing additional information to game design, MDAs, goals, or different levels of control functionalities. Therefore, the storyboard template used in class has been enhanced by these components. In future, more workshops are planned which emphasize "girls-only" environments, to reinforce especially girls with playful coding activities and help them in designing sophisticated games.

## Acknowledgements

This work has been partially funded by the EC H2020 Innovation Action No One Left Behind, Grant Agreement No. 645215.